\documentstyle[12pt]{article}
\begin{document}
\vskip 50 mm
\begin{center}
\bf{ THE NON-ORGANIC THEORY OF THE GENESIS OF PETROLEUM } 
\end{center}
\vskip 20 mm
\begin{center}
{\bf Samar Abbas } \\
Dept. of Physics ,Utkal University \\
Bhubaneswar-751004, India \\
\vskip 5 mm
and \\
\vskip 5 mm
Institute of Physics \\
Bhubaneswar-751005, India $^1$\\
\vskip 10 mm
e-mail : abbas@iopb.ernet.in \\

\vskip 20 mm
{\bf Abstract }

\end{center}
Recent advances in interdisciplinary fields as diverse as
astrophysics, cosmogeophysics, nuclear geology, etc.
have led to interesting developments in the non-organic
theory of the genesis of petroleum. This 
theory, which holds that petroleum is of an abiogenic 
primordial origin, provides an explanation for certain 
features of petroleum geology
that are hard to explain
within the standard organic framework. If 
the non-organic theory is correct, then 
hydrocarbon reserves would be enormous and
almost inexhaustable. 

\vskip 30 mm
1. address for correspondence

\newpage

{\bf (1) INTRODUCTION }                   
\vskip 5 mm
Petroleum is the foundation of this Industrial Civilization. It 
is from petroleum that the world obtains its chemicals, its
fuel for automobiles, engines, airplanes, etc.
and its energy supply for its power stations.
Empires have risen and fallen
due to the annexation or loss of oil fields. Hence, the origin 
of petroleum and the assurance of future energy supplies is of 
the utmost importance if this world is to continue as it is. It 
is generally believed that
currently recoverable crude oil resrerves will be nearing
exhaustion within a few decades. This 
estimate is based on the conventionally accepted (organic)
theory of the origin of petroleum.  

However it does not necessarily follow that this civilization  
will fall into darkness. The origin of
petroleum still, despite the immense amount of research 
devoted to it, has more uncertainties concerning it than any
other common natural substance $^1$ .

There are two basic frameworks : the standard organic theory; 
and the nonorganic theory. The former 
holds that petroleum is of an organic origin
and is the 
currently favoured proposal. It predicts limited reserves
worldwide; moreover Indian reserves are predicted as minimal. The 
latter maintains that it
is of a non-organic genesis, supposedly of primordial 
origin. On 
the basis of this theory, oil 
resources would be much larger than those predicted by the 
biogenic theory. India, oil-poor in the biogenic framework, is 
predicted to be oil-rich in
the non-organic one.

Unfortunately the abiogenic theory and its implications are not 
well known. Moreover, both opposing sides have taken uncompromising,
even fundamentalist views on occasions. There 
is hence a crucial need, especially 
for nations such as India, to objectively assess the
situation and investigate the latter possibility more 
carefully; especially since, as 
we shall discuss below, the evidence in 
favour of either candidate is inconclusive and the question still
remains an open one.  

\newpage     
{\bf (2) THE ORGANIC THEORY }
\vskip 10 mm
{\bf (a) Outline }
\vskip 5 mm
The organic theory holds that the first stage of the genesis of
petroleum involves 
plankton (single-celled organisms that float on the oceans). These 
die and gradually accumulate on the ocean floor. Other sediments
start accumulating too, and after a few million years the plankton
are buried under several km of sediment. The plankton, which have
remained unoxidised, under the increased values of pressure and 
temperature, are now transformed into kerogen. Under favourable
conditions of time and temperature this kerogen, after 
further burial and heating, is transformed, via cracking, into 
petroleum and natural gas. These 
then migrate towards the surface and end up
either reaching it (and
drying up to yield bitumen or tar) or  
being arrested on the way in traps (where, millions 
of years later, drillers of the 
present industrial age make their big strikes)
$^2$ .
\vskip 10 mm
{\bf (b) Advantages }
\vskip 5 mm
Traditionally, the following points have been considered as
supporting the biological theory: 

(1) Since it is known that hydrocarbons can be produced by
photosynthesis, it is natural to expect petroleum to be
of an organic origin..

(2) Molecules thought to be of biological origin, e.g.:
porphyrins, isoprenoids, hopanoids, etc.  
were found in petroleum, thereby providing support for the
organic theory. 

(3) The organic carbon in plants is depleted in carbon-13 due to the
process of photosynthesis. In dead organic material the C-13 is 
further depleted due to radioactive decay. Since 
it was found that most petroleum and natural
gas showed the same depletion, it was viewed as a strong proof in 
favour of an organic origin.

(4) Sediments are the most important host rocks 
yielding petroleum, i.e. the 
oil produced from oil wells is generally obtained from 
a porous sandstone deep below. Often sediments are associated with
biological material that could have acted as a source of the
petroleum.

(5) The existence of large quantities of oil shale from which
a hydrocarbon mix similar to petroleum could be distilled  
was seen as a support in favour of an organic origin. This 
followed easily, since
the oil shale was taken to be the kerogen source rock which, on
sufficient burial, purportedly yielded petroleum.
\vskip 10 mm
{\bf (c) Disadvantages }
\vskip 5 mm
However the following observations go against the organic theory :

(1) The discovery that meteorites contain hydrocarbons came
as a great blow to advantages no. 1 and 2 of the organic 
theory. Porphyrins 
and isoprenoids have been found on meteorites $^3$ . In 
addition, the outer planets contain large amounts of 
hydrocarbon.

(2) The concentration of oil in the Middle East 
implies that that region must have been exceptionaly prolific in 
plant and animal life over long periods of the Earth's 
history. This is unlikely, since life tends to be more 
dispersed, even today.

(3) The biological supports of optical activity and an odd-even
 effect disappear 
 at low levels. There is a sharp cutoff to
 the effect of optical activity: Petroleum in Philippi's
 study $^4$ was found to be optically active if derived from a
 reservoir with a temperature below 66 C, but surprisingly petroleum 
 from deeper levels of the same field did not exhibit the phenomenon
 of optical activity. T. Gold proposed that a certain bacteria ceases
 action above 66 C $^5$ , but he unfortunately did not suggest any
 candidates. 

(4) Methane occurs in giant ocean rifts, in continental rifts and 
  the lakes that occur nearby, e.g. dissolved in the waters of
  the East African Lake Kivu $^6$ ,
  as methane hydrates in
  permafrost, in active volcanic and mud volcanic regions, as 
  well as at great depths of more than 10 km as geopressured gas
  etc. A biological origin for this methane can be virtually ruled 
  out.

In light of these difficulties one should consider the other rival
nonorganic theories as serious possibilities. They forecast
much larger oil reserves than previously imagined and that too
in regions which, according to the organic theory, should be 
devoid of 
all petroleum.

\newpage
{\bf (3) THE NON-ORGANIC THEORY }
\vskip 5 mm
{\bf (a) Historical Development }		
\vskip 5 mm		

The Non-organic theory of the genesis of petroleum has a
long history, dating back to the early days of the oil
industry. Its development has led to the birth of
a number of variants, the most 
important of which are outlined below :

\vskip 5 mm
{\it (i) Metal Carbide Theory }
\vskip 5 mm
The founder of the non-organic theory was Mendeleev, the 
Russian chemist who proposed the modern version of the
periodic table. In 1877, he 
wrote that the petroleum deposits of the 
world seem to be controlled more 
by large-scale tectonic features than by the ages of sedimentary
rocks $^7$ . To explain these observations he put forth
the metal carbide theory. Many contemporary 
investigators, mostly Russian, supported
Mendeleev's view.

In this model metal carbides deep within the earth
reacted with water at high temperatures to form acetylene
which subsequently condensed to form heavier hydrocarbons (as
is readily formed in the lab). The following reaction
        
		$CaC_2$ + 2$H_2O$ = $C_2H_2$ + $Ca(OH)_2$
		
is still popular amongst some astronomers 
and certain Russian geologists
as a major petroleum-forming possibility $^{2 , p.271} $.

\newpage

{\it (ii) Nebular Condensation Theory }
\vskip 5 mm
In 1890 Sokoloff proposed that `bitumina', at that time meaning
the whole range of hydrocarbons from petroleum to 
tar, precipitated as rain 
onto the newly forming earth from the original nebular matter 
from which the Solar System was formed. In modern terminology
he simply suggested that petroleum originated from 
meteorites. Later, he claimed, this petroleum was
ejected from the earth's interior into the surface 
sediments $^8$ . Recently 
this idea has been supported by F. Hoyle, who
proposes that not only oil, but life itself has extra-terrestrial
origins $^9$ .
\vskip 5 mm
{\it (iii) Volcanic Origin Theory }
\vskip 5 mm
This postulate involves outgassing of the mantle via volcanic 
activity $^{10}$ .
\vskip 5 mm
{\it (iv) Earthquake Outgassing }
\vskip 5 mm
This theory proposes that outgassing occurs via deep 
faults, and that this is still
occurring today. The 
detailed mechanism has a long history. V.I.Vernadsky 
( 1933 ) propounded the 
notion that hydrocarbon compounds would be stable against
dissociation and oxidation at great depths and would replace
carbon dioxide as the chief carbon-bearing fluid.
N.A.Kudryavtsev ( 1959 ) set forth the observations 
supporting what was
later to be known as Kudryavtsev's rule (to be discussed shortly).

T.Gold $^{5 , 11}$  has become the main proponent of the idea 
of a non-organic origin in the West. Due to
his initiative, a hole was drilled into crystalline basement rock
which Gold predicted should yield petroleum. Only noncommercial
quantities of petroleum, if any, were found. Moreover most of
Gold 's colleagues were not convinced.
However, recently a nearby hole did strike oil (see below).
This is now the most commonly known
variant. 

However,since Western nations either possess or control much of the
world's petroleum reserves, there is no incentive to innovate.
Since the nonorganic theory predicts petroleum in much larger
quantities and in areas hitherto considered unfavourable,
it is petroleum-poor countries like India who stand most to 
gain. Hence ,it is they who should take the risks of exploring the
non-organic theory, both theoretically and experimentally.

These are the prime variants of the theory. From now on, the word
`non-organic theory ' shall be taken to propound 
merely a primordial ( i. e. dating from the birth of the Earth )
origin of petroleum which has been migrating outwards from 
great depths of the Earth to form all hydrocarbon deposits
from tar and tar sands to oil shale. The 
detailed mechanisms mentioned above shall 
not complicate the issue during the course of the following
discussion. 
\vskip 10 mm
{\bf (b) Outline of the Theory }
\vskip 5 mm
The theory suggests that most of the hydrocarbons on earth are
in fact primordial. Carbonaceous chondrites appear to have been the
most abundant source rock during the formation of the earth. This
type of meteorite contains a significant amount of hydrocarbons.
As the earth formed, it would have acquired these hydrocarbons 
via accretion (bodies of roughly equal size clumping together 
through collisions), and later through meteorite impacts
(including hydrocarbons formed by the reaction of meteoritic
carbon with $H_2$ at high pressures and temperatures on
impact). Then 
as the earth gradually cooled, a solid crust developed, while the 
interior remained liquid or semisolid. The volatile substances
would be expelled from the interior. It is such gases that 
yielded, after biological modification, the present atmosphere.
That hydrocarbons are being evolved from the inner parts of the 
earth is evident from the presence of mud volcanoes, flames
seen during earthquakes, etc. On the way up, it is supposed,
the oil ( dissolved in methane ) 
would be trapped in suitable formations creating 
the world's oil and gas,
tar sands, oil shales, bitumen, mud volcanoes, etc $^{11}$  . 
Kropotkin and Valyaev pointed out that the hydrocarbons, carried
upwards by streams of compressed gases, would have two possible 
destinies : 1. In volcanic regions, they would be oxidised to 
carbon dioxide and water, and  2. In `cool' regions the 
hydrocarbons would form oil and gas reservoirs after condensation
from the rising stream at levels possessing the requisite values
of temperature and pressure $^{12} $.

\vskip 5 mm
{\bf (c) Evidence in favour of the Non-Organic Theory }
\vskip 5 mm
{\it (i) Geographical Location }
\vskip 5 mm
The major oil fields of the world are concentrated on or near belts
of major tectonic activity or in fact along fault zones. Some of the
phenomenal Arabian fields, the world's largest petroleum province,
lie along the Persian Zagros Mt. belt. The large North Sea reserves 
that have made much of Northern Europe self sufficient in oil
production lie along the North Sea trench. The oil fields of 
Indonesia and Burma closely follow the seismic belt running from
New Guinea to Burma, while the oil fields of Gujarat appear to be
associated with the Cambay fault. Hydrocarbons 
are found in the Red Sea rift 
Valley, the East African rift and the eastern branch of the Pacific
Rift. These and many other examples that exist should illustrate the
association of hydrocarbons with large deep-seated cracks in the
Earth's crust rather than any local sediments. However, note
that the idea of deep-seated cracks may also be required
to explain the migration of petroleum within the organic theory.

According to the non-organic theory, petroleum should 
occur universally in areas of tectonic activity. This does not
appear to hold true, and this seems to be a problem for the
non-organic theory.

\vskip 5 mm  
{\it (ii) Multilevel Fields }
\vskip 5 mm
It is observed that petroleum, in at least small quantities ,is
often present in horizons below many accumulations, largely 
independant of the composition and mode of formation of the
horizon. This is known as Kudryavtsev's 
rule, and several
examples of it have been noted $^{ 13 }$ . The suggestion
that the petroleum seeps from underneath is supported by the
evidence of fractionation, although this can also be explained
by migration from deep source rocks within the organic 
framework as well.

Methane-bearing strata in the same column show a progressive
depletion in the isotope carbon-13 as one rises 
from lower levels to higher levels $^{ 14 }$. 
The organic theory holds that petroleum originating
from source rocks buried several km within the earth
explains these properties. The 
oil and gas formed would migrate upwards, thereby explaining
both fractionation and
Kudryavtsev's rule. However, this effect would be a natural
consequence of the upward migration of primordial gases,
with the heavier isotopes of carbon 
rising more slowly than the lighter one.

\newpage
{\it (iii) Stability with Depth }
\vskip 5 mm 
It was once thought that petroleum and natural gas would not be able
to survive at depths greater than a few tens of kilometres below the
surface as the temperatures occuring there exceeded those observed 
to destroy petroleum and natural gas in labs. Hence, it 
was reasoned, it 
was pointless to look for either the fuel or its origin in the
depths of the Earth.
However that picture has changecd radically. Huge quantities of
gas have been discovered at great depths e.g. in the Anadarko basin
in Oklahoma.Reservoirs of `geopressured gas' have been found to
underlie all major oil bearing regions. These are sandstones 
and shales containing enormous amounts of gas dissolved in 
salt brines. Reserves of such gas are estimated at 60000 TCF
(trillion cubic feet )in the U.S. alone $^{ 15 }$, 
exceeding by several factors the total conventional gas
reserves of the world.

In addition vast domains
of gas exist in open fractures of non-sedimentary basement 
rock. A deep hole 
currently being drilled in Germany has found these at
depths of up to 4km $^{ 16 }$ . Theory 
had to be revised, and Chekaliuk $^{ 17 }$ showed that not only
could natural gas exist at extreme depths but petroleum could, too.
The earlier experiments had simply not been done at the correct 
pressures. In fact, the pressures encountered stabilize oil and
natural gas against dissociation despite extreme temperatures, so
that methane could exist up to 30km with only 5 $\%$
dissociated. Further 
thermodynamic calculations show that petroleum
itself is mostly stable between 30 to 300 km.
Although this is heartening from the perspective of the non-organic
theory, this does not necessarily go against the
organic theory, as we discussed earlier. 

Sugisaki and Nagamine have recently investigated the 
thermodynamic equilibrium of light hydrocarbon gases $^{ 18} $.
Thermodynamic equilibrium of a gas is revealed by the 
concentrations of its constituents. They studied the reaction

             $CH_4$ + $C_3H_8$ = 2$C_2H_6$ 

At equilibrium the graph of [$CH_4$]x[$C_3H_8$] vs. [$C_2H_6$]
is a graph of constant slope. Moreover, the temperature of 
equilibrium can be calculated from this graph. It was found 
that hydrocarbon gases 
released by crushing plutonic rock and
natural gas from deep wells displayed these features, indicating
chemical equilibrium. On the other hand, gases issuing from a peat 
bog, shallow gas wells, and, significantly, pyrolysis products from
kerogen, coal and other organic substances did not  
, although the temperature was 350 C (thus exceeding the 
equilibrium temperature of 180 C for the plutonic gases)
and the longest experimental
period was 555 days. Quite naturally this is a puzzle , 
since if kerogen were the source of petroleum, then the 
hydrocarbons released through pyrolysis of kerogen should
display the chemical equilibrium shown by the
deep-level hydrocarbons. It should be noted that this work,$^{ 18 }$
pertains to thermodynamic equilibrium of light hydrocarbon gases and
not to the stability of the same.

The team explained these 
observations in terms of the organic theory as follows:
After the decomposion of petroleum, kerogen and other
heavy hydrocarbons, the gases attain chemical equilibrium at the
high temperatures existing at great depths. As the gas cools,
either by upward migration or by cooling of the volcanic rock 
(in the case of the plutons), the gas composition is frozen in 
once the temperature becomes so low that the gas composition 
effectively freezes in. To explain the negative result, i.e. the
lack of equilibrium of 
the pyrolysis products, they are forced to make the
rather unilluminating assumption that the rate of reaction
was so slow that equilibrium could not be attained even after 
1$ 1 \over 2$ years.

A far more natural explanation, which the authors cited above 
chose not to 
investigate, is using the non-organic theory. If it is assumed 
that the hydrocarbons are primordial and originate from the depths of
the earth, then they would naturally display the signatures of 
chemical equilibrium as it would have been subject to high 
temperatures for a much longer time. As the hydrocarbons
migrate upwards, they would, at shallow levels, be 
invaded by bacteria
thereby losing the signatures of equilibrium.

Critics suggest that oxidation would destroy any petroleum anyway
turning the hydrocarbons into $CO_2$ , $H_2O$ and coke, 
the constituents of 
volcanic gases. Since these 
come from deep inside the Earth, the 
primordial hydrocarbons, even if they existed, would have 
been destroyed. However that is not the full story. Substantial 
evidence indicates that unoxidised carbon exists at great depths.
Unoxidised carbon can also exist at great depths 
within the organic framework, coke being produced by the
decomposition of hydrocarbons.

\newpage
{\bf (d) The Existence of Unoxidised Carbon at Great Depths }
\vskip 5 mm
The following evidence suggests that large masses of unoxidised
carbon exist at great depths:
\vskip 5 mm
{\it (i) Diamonds }
\vskip 5 mm
Diamonds provide evidence of 
the existence of carbon at 
great depths in an unoxidised form, i.e. other 
than $CO_2$, since diamond is pure carbon.

The diamonds might have been associated with extremely violent
explosions, since if the 
diamonds had risen slowly along with magma of the type spewed
out by conventional volcanoes (e.g. Hawaii), they would have been
transformed into graphite. As thermodynamical calculations and 
experience in the production of artificial diamonds show, diamonds
must be cooled quickly to exist. Moreover the associated host rock
contains other high pressure minerals like peridotite, (believed 
to be one of the principal 
constituents of the mantle), etc. The primary
deposits of diamonds are the rare kimberlite 
pipes, named 
after the now legendary South African town of Kimberley which
was once the chief source of these stones. These are 
deep vertical shafts, funnel-like 
near the surface and gradually narrowing with 
depth, presumably 
becoming a fissure extending all the way to the upper
mantle where the pressure and temperature are suitable for the 
formation of diamonds (45 kbars and 1000 C approx.). Why are
they so deep? How were they formed?Moreover ,if they are volcanic in
origin, as they appear to be, why is no lava associated with 
them? The only plausible conclusion is that the pipes were caused by
extremely violent eruptions of gas that blasted a hole through 150km
of overlying dense rock. The observation that the pore spaces of 
natural diamonds contain highly compressed gases including $CO_2$
and $CH_4$ $^{ 19 }$  and the result that heavy hydrocarbons
clearly distinguishable from the surrounding rocks exist in the East 
Siberian pipes $^{ 20 }$ , including the observation that 
bore holes into
these pipes yield significant quantities of $CH_4$ $^{ 21 }$
enforce the conclusion that 

(1)Unoxidised carbon exists in the outer mantle in the form of 
methane and hydrocarbons, as does pure carbon.

\newpage

(2)Volatile-rich regions exist in the inhomogeneous mantle which
have been giving off hydrocarbon
gases long after the formation of the earth's
crust which can build up such great pressures that they simply crack 
through the crust in violent explosions. 

\vskip 5 mm
{\it (ii) Earthquakes }
\vskip 5 mm
The eruptions of gas mentioned above, the generation of
which is not disputed by the organic theory, should 
cause earthquakes. In 
fact, earthquakes have ben observed to be associated with gas
ejection throughout recorded history :
\vskip 3 mm
{\sf (a) Greco-Roman civilization } :
\vskip 2 mm
Anaxagoras first proposed the theory that gases ('air') were the
cause of earthquakes $^{ 5, p.49} $. Seneca,Pliny,Pausanias and
Aelian mention the evolution of `wind', strange animal behaviour,
great flames rising from the ground, loud roaring noises, foul
smell and peculiar fog and the development of peculiar 
odour and muddy appearance in the water of wells and springs
occuring several days or months prior to the earthquake. 
\vskip 2 mm
{\sf (b) Anglo-Saxon Age } :
\vskip 2 mm
Newton wrote that he felt `sulphuurous fumes'were the cause of
earthquakes $^{ 5, p.51 }$ . Mitchell (1761) 
reasoned that gases caused the 
slow,visible oceanlike waves that roll across landscapes
during major earthquakes. He also points out that the
sudden deaths of large numbers of fish would be most naturally 
explained by the evolution of poisonous gases from vents in the
ocean floor.
Alexander von Humboldt (1822) summarized the then accepted theory 
as `elastic fluids seeking an outlet to diffuse themselves into the
atmosphere'being the cause of earthquakes. Thousands
of fish, many of a nature previously unknown to local fishermen,
were found floating on the water in Monterrey Bay on the day of 
the destructive San Fransisco earthquake of 18 April 1906 
$^{ 5, p.63 }$.
Similar reports come from Japan. Hydrogen Sulphide, highly toxic to 
fish, is a likely candidate, killing the bottom dwelling fish 
that are not normally caught.
\vskip 2 mm
{\sf (c) Chinese civilization } :
\vskip 2 mm
More recently, at the Sungpan-Pingwu easrthquake (Aug. 1976),
outbursts of natural gas from rock fissures were reported.
Moreover, these sometimes ignited, creating fireballs.
A total of 1000 were sighted. A few hours before the earthquake, 
the water in local wells was observed to exhibit a violent 
bubbling $^{ 5, p.61-62} $.

These reports spanning recorded history show that methane gas is
closely associated with earthquakes. In fact it is not
unreasonable to suggest that earthquakes are caused by enormous
build-ups of highly compressed gases containing mostly methane.
This is in fact a strong support in favour of the non-organic theory
, since the amounts of methane evolved are too large to have been 
produced by biological sources.
\vskip 5 mm
{\it (iii) Mud Volcanoes }
\vskip 5 mm 
Mud volcanoes are volcanoes that, instead of ejecting lava and gas
like ordinary lava volcanoes, emit mud and gas instead. The cones
built up by them, consisting of solidified mud, are similar 
to, but
smaller than, those built by the lava volcanoes. They emit mostly
methane, while smaller amounts of other hydrocarbons are also present,
including other inorganic gases like He, $H_2$, $CO_2$ or steam.
Many mud volcanoes simply eject high pressured unconsolidated mud.
In contrast, lava volcanoes emit mostly carbon dioxide and water.
Mud volcanoes closely follow the underlying fault lines. 
This is not
just commomplace, but holds for all mud volcano regions of the 
world. Moreover, the 
quantities of gas required to produce the Soviet 
mud volcano fields have been estimated to be several times the 
total gas content of the largest known gas field $^{ 5, p.101 } $.
How does one explain this?

The conventional explanation in terms of the organic framework
is that the gas is generated by bacterial action on the organic 
content of the mud. However, this has some problems: 

(1)The gas so generated would bubble up on a continuous basis, and
hence extremely violent explosions of the type observed in the
major mud volcano regions of the world would be extremely unlikely,
as large concentrations of gas most probably not build up.

(2)Chemical analysis reveals that the methane also contains
significant amounts of ethane,propane and other hydrocarbons.
Moreover, mercury, helium and other trace elements occur in the
gases. 
The carbon isotope ratio is sometimes quite
different from that expected to be obtained from a biologically
derived material.

The inorganic explanation is that the gradually upward migrating
gaseous hydrocarbons build up beneath impervious rocks, and then
after having built up sufficient pressure, smash through the 
overlying rock, creating violent explosions of the type observed.
These violent displacements of gas will cause violent turbulence
of the water, which would stir up the fine-grained sediment,
creating the mud so characteristic of these volcanoes.
\vskip 5 mm
{\it (iv) Pockmark-like Craters on the Ocean Floor }
\vskip 5 mm
Crater-like markings on the ocean floor have been reported from
the Adriatic, the North Sea, the Gulf of Mexico, the 
Orinoco Delta,
the South China Sea, the Baltic, the Aegean, near 
New Zealand, and
off Nova Scotia $^{ 22} $. Sonar experiments in the North Sea reveal
shallow, circular ridges ranging from a few metres to 200 metres in
diameter over an area of 20000 square kilometres, roughly 
coinciding 
with the oil and gas producing region. It appears that individual 
events were responsible for creating large fields of these 
`pockmarks', since one set of pockmarks occurs 10m below an 
overburden of more recent sediment, while the other is visible on the
surface. Hence it is estimated that while one 
such event occurred within the
last thousand years, another occurred 10000 years ago. Since small
trickles of gas produce small steep-sided cones of mud (as in the
Gulf of Mexico, where bubbles issue from the top of these 
miniature volcanoes), sudden releases of gas must be responsible 
for the craters. Primordial gas is a good candidate to explain 
the pockmarks. 

The author also points out a remarkable coincidence between the
major mud volcano regions of the world and the major oil-producing
areas :

1. The Persian Gulf

1. The Caspian
  
3. Indonesia

4. Venezuela

The South Alaskan mud volcanoes emit mostly carbon dioxide and are
situated near lava volcanoes. Only 3 mud volcano regions are not
correlated with any oil-producing regions : S. Italy, N Zealand,
Black Sea. This connection arises naturally in the non-organic
approach, since mud volcanoes indicate cool regions of hydrocarbon
migration.

\vskip 10 mm
{\bf (e) Extra-terrestrial Hydrocarbons }
\vskip 5 mm
{\it (i) Meteorites }
\vskip 5 mm
If primordial hydrocarbons were incorporated in the earth during the
process of formation, then one should expect to find such substances
in ancient material dating from the formation of the solar system. 
Such material exists in the form of carbonaceous chondrites, 
a class
of meteorites. Moreover, this type of meteorite seems to be very
common. In fact, asteroids and interplanetary material seem to
be of largely carbonaceous derivation $^{ 23} $.

Materials previosly thought to be exclusively biological in origin
have now been found on meteorites. 
Porphyrin-type molecules are found in meteorites and are almost 
certainly not of a biological derivation $^{ 3} $.

\vskip 5 mm
{\it (ii) Planets }
\vskip 5 mm
The outer planets have their atmospheres largely in the form of 
hydrocarbons, chiefly methane.
Uranus' atmosphere may contain as much as 14$\%$ of methane gas 
$^{ 24, p.221 }$. Neptune's atmosphere consists of hydrogen, helium and
methane while the inner liquid shell is thought to consist of 
water, methane and ammonia $^{ 24, p.233} $.
\vskip 5 mm
{\it (iii) Comets }
\vskip 5 mm
Halley's comet (1986) was found to emit hydrocarbon gases. The core
was observed to be black, presumably because of it being
composed of carbonaceous material. Lang and Whitney describe the  
interior as blacker than coal ,its blackness perhaps being due to 
`an admixture of minerals, organic compounds and metals'
$^{ 24, p.254 }$.

\vskip 5 mm
{\it (iv) Isotope and Trace Element Anomalies }
\vskip 5 mm
The following peculiarities point to an extensive upward
migration of deep fluids. Moreover, mantle-derived material 
occurs in association with petroleum :

(a) $\cal HELIUM $

The noble gas helium is closely associated with petroliferous 
regions. In fact the world obtains its commercial supply of 
helium by separation from natural gas $^{ 11, p.69} $.

(b) $\cal ARGON $-40

The noble gas argon and its isotope Ar-40 occur in extraordinarily
high concentrations in gas fields $^{ 11,p.70 }$.
Moreover, assuming the source 
of the 0.1$\%$ Ar by volume in the huge Panhandle gas field to
be the source rock itself implies that the source 
rock must have been 100$\%$ 
potassium to supply the required levels of Ar-40 $^{ 29} $.
Moreover, high values of 
Ar-40/Ar-36 are taken as an indication of mantle-derived material
$^{ 29, p.417} $ ; strikingly petroleum displays this signature.

\vskip 5 mm
{\bf (f) Experimental Verification }
\vskip 5 mm
The final proof would involve an actual experimental verification 
of the theory.
Deep wells are good tests, since organic materials cannot occur in 
crystalline basement rocks. Several are under way :
\vskip 5 mm
{\it The Kola Superdeep Hole }
\vskip 5 mm
At 12 km, this is the world's deepest well ( 1984 ). Located 
in the Kola
peninsula, now Russia, it reached deep down into the crystalline 
basement. The drilling released flows of gas at all levels.
The liberated gases included helium, hydrogen, nitrogen,
methane and other hydrocarbons and $CO_2$ $^{ 30} $ .
This provides convincing support for the suggestion that 
hydrocarbon gases exist at such great depths inside the earth 
that they cannot be of a biological origin.

\vskip 5 mm
{\it The German Deep Hole }
\vskip 5 mm
This hole is located in Windischeschenbach ( Oberpfalz )
The depth reached during the pilot drilling programme was 4 km 
( 1990 ).
From 3.2 km down the drill
encountered increasingly common cases of highly concentrated 
salt brines with gases like methane and helium in open caves in
the rock $^{ 16} $ , but no petroleum . 

\newpage
{\it The Swedish Hole }
\vskip 5 mm
The discovery of an oil and/or gas field in a
location ruled out by the organic theory would settle the matter
once and for all. Hence, T. Gold $^{ 5,p.172}$, after studying 
various formations across the world, concluded that the Siljan 
Ring, Sweden was the best candidate for the job.
This is the largest impact crater
in Europe. According to the non-organic theory the 
impact could have led to the formation of sizeable hydrocarbon
deposits since the fractures created by the impact
would favour the upward migration of primordial
hydrocarbons. 
Although the field was located primarily on the 
basis of seismic data, 
numerous oil seeps have been noted in the small sedimentary 
deposits of the ring-shaped depression marking the crater,
carbonates characteristic of oxidised methane occur in the area
and seismic observations reveal zones of porosity stacked on top of 
one another. Hence the primary indications were favourable.

Finally , T. Gold succeeded in convincing investors and  
a project
began to prospect for petroleum in the area. This was largely
supported by the state-owned Swedish electricity utility
Vattenfall.
Drilling began in 1986. By late 1987 a depth of 6.5 km
had been attained, but no large commercial deposits
had been discovered  $^{ 25} $ .
Opponents saw in this the death knell of the non-organic
theory ( claiming the 
hydrocarbons detected were from the lubricating drilling mud
injected into the ground during drilling ), while T. Gold
proclaimed a victory ( claiming that significant amounts of
hydrocarbons were discovered, and that large amounts lay beneath )
$^{ 26} $.
Due to drilling difficulties, the project stopped short of its
target. It can be said that this did not rule in favour of either 
proposition. 

However, in 1989 the Swedish drilling company Dala Djupgas 
Produktions
recovered a small quantity of oil from 
6.7 km below the Siljan Ring.
Again critics 
dismissed the find as being recycled drilling fluid. The tables
were turned yet again when the same co. discovered oil in 1991
even at the shallow depth of 2.8 km at a nearby well, the
horizon of the petroleum being basement granite $^{ 27} $ . 
Moreover the previous 
objection was nullified as the drilling fluid used in this case 
was water only. The proponents of the organic theory claimed that
this oil was merely oil that had seeped into underlying fissures in 
the basement rock from oil shales, since the petroleum found 
in the oil shales and that in the basement rock were chemically very 
similar. However, the non-organic theory explains this as being due
to the upward migration of primordial petroleum; the two oils are
similar because their common source is the same. The upward
migrating hydrocarbons would have produced both the deep oil
and the oil shales, the shale providing a good trap rock
that could absorb the oil on its way up. Hence, the case
appears to have recently swung in favour of the non-organists.

Only drilling in the future by men 
of the calibre of Col. Drake ( the discoverer of the world's
first oil-field ), Dad Joiner ( the discoverer of the giant E. Texas
oil-field ) and 
P. Higgins ( the discoverer of the Spindletop oil-field )  
can yield the answer to this intriguing question.
\vskip 3 mm
{\bf (4) CONCLUSION }
\vskip 2 mm
The positive and negative features of the classical organic theory
have been discussed. This has been the traditionally accepted 
proposal, much work having been done in this field. The rival
non-organic theory has so far not been accepted due to the 
successes of the biological theory to date in elucidating 
certain properties of oil-fields. However, new results from deep
holes all across the world are difficult to describe in terms of
the biological theory. It has been shown that these new 
observations can be naturally explained within the non-organic 
framework, and that the older biological supports (mainly relating
to the presence of supposedly biological material in petroleum)
can also be incorporated. Hence a duplex theory combining features
of both theories may be the final victor. This would 
perhaps involve the enrichment of existing organic hydrocarbon
deposits through non organic hydrocarbons.

The abiogenic theory derives much of its
support from diverse and exotic fields such as astrophysics, 
cosmogeophysics, thermodynamics, nuclear geology, etc, and 
considerable
strides in the comprehension of these fields has led to a 
impressive growth of information in support of the non-organic 
theory. 
 
\vskip 3 mm
ACKNOWLEDGEMENTS
\vskip 3 mm
The author would like to thank Prof. S. Mohanty
and Prof. A. Abbas for  fruitful discussions
and many kind suggestions.
The author thanks the referees for helpful comments.

This work is supported through a fellowship of the 
University Grants Commision, New Delhi, India.

\newpage

{\bf REFERENCES }
\vskip 10 mm
1.  Hedberg, H. D., `Geologic aspects of origin of petroleum',
    Amer. Assoc. Petrol. Geol. Bull. {\bf 48} (1964) 1755-1803

2.  Encyclopedia of Physical Science and Technology, ed. R. A.
    Meyers, Academic Press (Orlando, Florida U.S.A.) 1987
    Vol. 10 p.269-280

3.  Hodgson, G. W. and Baker, B.L..`Porphyrins in meteorites:
    Metal complexes in Orgueil, Murray, Cold Bokkeveld and
    Mokoia-carbonaceous chondrites ', Geochim. Cosmochim. Acta
    {\bf 33} (1969) 943-958

4.  Philippi, G. T. `On the depth, time and mechanism of origin of 
    the heavy to medium-gravity naphthenic crude oils' , Geochim.
    Cosmochim. Acta {\bf 41} (1977) 33-52
 
5.  Gold, T. , `Power from the Earth', J.M.Dent and Sons Ltd. 
    London 1987 

6.  Deuser W. G. et al. `Methane in Lake Kivu: new data bearing on 
    its origin ', Science {\bf 181} (1973) 51-54

7.  Mendeleev, D. `L'origine du petrole ', Revue Scientifique, 2e Ser.
    ,VIII, 409-416

8.  Sokoloff, W. `Kosmischer Ursprung der Bituminas', Bull. Soc.
    Imp. Natural Moscau, Nuov. Ser. 3 (1889) 720-739

9.  Hoyle, F.and Wickramasinghe, C. , Newscientist {\bf 76} (17 
    Nov.1977) p.402 , as well as   
	 
	Newscientist {\bf 77} (19 Jan.1978) p.139 

    Nature {\bf 266}  (1977) 241-243
	
	Nature {\bf 270}  (1977) 323-324

10. North, F. K. `Petroleum Geology', Allen and
    Unwin Inc. 1985 p.37-42

11. Gold ,T. `The origin of natural gas and petroleum, and the 
    prognosis for future supplies', Ann. Rev. En. {\bf 10} (1985)
	53-77

12. Kropotkin, P.N. ,and Valyaev, B.M. `Development of a theory of
    deep-seated ( inorganic and mixed ) origin of hydrocarbons',
	Goryuchie Iskopaemye : Problemy Geologii i Geokhimii Naftidov
	i Bituminoznykh Porod \\
	(N.B.Vassoevich, ed.), pp.133-144.
	Akad. Nauk SSSR. 	

13. Kropotkin , P.N., and Valyaev, B.M. `Tectonic control of Earth
    outgassing and the origin of hydrocarbons', Proc. 27th Intern.
	Geol. Congr. {\bf 13} (1984) 395-412 VNU Science Press.	

14. Galimov, E.M. `Isotopic Composition of carbon in gases of the 
    Earth's Crust ', Internat. Geol. Rev. {\bf 11} (1969) 1092-1103 

15. National Geographic , `Natural Gas- The Search goes on'
    {\bf 154} ( Nov. 1978 ) p.632-p.651

16. BMFT ( Bundes-Ministerium fuer Forschung und Technologie )
    -Journal ( Journal of the German Ministry of Research and 
	Technology ) Nr.3 June 1990 p.13

17. Chekaliuk, E.B. `The thermal stability of hydrocarbon systems in
    geothermodynamic systems ', Degazatsiia Zemli i Geotektonika
	(P.N.Kropotkin, ed.), pp.267-272

18. R. Sugisaki and K. Nagamine ,
    `Evoluton of light hydrocarbon gases
    in subsurface processes :Constraints from chemical equilibrium ',
	Earth and Planetary Science Letters {\bf 133} (1995) 151-161 

19. Melton, C.E., and Giardini, A.A. `The composition and 
    significance of gas released from natural diamonds from
	Africa and Brazil', Amer. Mineralogist {\bf 59} (1974) 775-782

20. Kravtsov,A.I. et al `Distribution of gas-oil-bitumen shows in the
    Yakutian diamond province',Int. Geol. Rev. {\bf 27} (1985) 
	1261-1275

21. Kravtsov, A.I., et al.`Gases and bitumens in rocks of the 
    Udachnaya pipe ', 
	Dokl. Akad. Nauk SSSR, Earth Sci. Sect.
	{\bf 228} (1976) 231-234

22. Hovland, M., et al .`Characteristic features of pockmarks on the
    North Sea floor and Scotian Shelf', Sedimentology {\bf 31} (1984) 
	471-480

23. Chapman, C.R.,`The nature of asteroids ', Scientific American
    {\bf 232} ( Jan.1975 ) 24-33

24. K. R. Lang and C. A. Whitney , `Wanderers in Space',
    Cambridge University Press 1991

25. Newscientist {\bf 115} ( 24 Sept.1987 ) p.26 

26. Gold , T. , Newscientist {\bf 116} ( 15 October 1987 ) p.67

27. Aldhous, P. , Nature  Vol. {\bf 353} ( 1991 ) p.593

28. Ozima, M. , `Noble Gas State in the Mantle ',
    Rev. of Geophysics {\bf 32} ( November 1994 ) 405-426

29. Pierce, A.P. et al `Uranium and Helium in the Panhandle gas 
    field, Texas and adjacent areas', US Geol. Surv. Prof. Paper
	454-G
	
30. Kozlovsky, Ye. A. `The world's deepest well', Scientific
    American {\bf 251} ( no. 6, 1984 ) 98-104

\end{document}